# On the Base Station Selection and Base Station Sharing in Self-Configuring Networks

S.M. Perlaza, E.V. Belmega, S. Lasaulce,  and M. Debbah,

### Abstract

We model the interaction of several radio devices aiming to obtain wireless connectivity by using a set of base stations (BS) as a non-cooperative game. Each radio device aims to maximize its own spectral efficiency (SE) in two different scenarios: First, we let each player to use a unique BS (BS selection) and second, we let them to simultaneously use several BSs (BS Sharing). In both cases, we show that the resulting game is an exact potential game. We found that the BS selection game posses multiple Nash equilibria (NE) while the BS sharing game posses a unique one. We provide fully decentralized algorithms which always converge to a NE in both games. We analyze the price of anarchy and the price of stability for the case of BS selection. Finally, we observed that depending on the number of transmitters, the BS selection technique might provide a better global performance (network spectral efficiency) than BS sharing, which suggests the existence of a Braess type paradox.

## I. Introduction

In this paper, we consider the case where several radio devices aim to obtain wireless connectivity by using several base stations (BS). Here, each device must strategically determine the set of BSs to use, as well as the corresponding power level allocated to each BS to maximize its own spectral efficiency in bps/Hz. In this context, we consider two different scenarios. First, we let each player to use a unique BS (BS selection) and second, we let them to simultaneously use several BSs (BS Sharing).

Note that if only one BS is considered, our model simplifies to a multiple access channel (MAC). Here, when all transmitters access the BS using the same carrier, each device uses its maximum transmit power. When, the BS is accessible through out several carriers, each transmitter uses a water-filling power allocation (PA) considering the observed multiple access interference as background noise at each carrier [1]. In the first case, such solution is Pareto optimal, if and only if the sum of the achieved Shannon rates lies in the convex hulk of the capacity region of the corresponding MAC [2]. Generally, this condition may require certain coordination between the transmitters, which can be achieved by using pricing methods [3]. Conversely, in the second case, the solution is always Pareto optimal [1]. In a more general context, when there exists several BSs and regardless of the performance metric, the model remains being a subject of intensive research [4], [5], [6], [7], [8], [9].

Up to the knowledge of the authors, the state of art of the BS sharing and BS selection scenarios is described by the following contributions: [10], [4], [5], [8]. In [10], the BS selection problem is investigated by considering that each node is characterized by a fixed single user spectral efficiency. Here, the authors showed that based on the scheme of exponential learning, players converge to an evolutionarily stable equilibrium. Additionally, the authors showed that the price of anarchy of such a game is unaffected by disparities in the nodes' characteristics. In [8], the authors studied the BS selection scenario assuming that the transmitters aim to minimize their transmit power level required to achieve a target signal to interference plus noise ratio (SINR). Here, the interaction between the radio devices is modeled as an atomic and non-atomic potential game [11] to study the existence, uniqueness and efficiency of the NE. Other contributions using potential games for radio resource allocation are [12], [13], [14], [15], [16]. In [5], the non-atomic extension of the BS selection game and the atomic extension of the BS sharing game are investigated. Therein, the performance metric is the Shannon rate and channel realizations are considered identical for all transmitters. Regardless of the possibly unrealistic assumption, the authors of [5] identified the existence of at least one NE in the non-atomic BS selection game and the existence of a unique equilibrium in the atomic BS sharing game. In [4], the authors study the BS selection and sharing scenarios when the number of receivers is equal to the number of transmitters and the performance metric is their overall SINR, i.e., the sum of the SINRs obtained at each BS. In this context, it is showed that when all players observed the same channel realization (as in [5]), restricting each player to choose only one BS produces a socially optimal NE. Conversely, when the players are left free to share their powers among several BSs, their utilities are strongly decreased. This effect is known as the Braess Paradox in the frame of congestion games [17].

In this paper, we tackle the BS selection and BS Sharing scenarios by modeling them as potential games. Contrary to previous works [8], we consider as performance metric the spectral efficiency of each player and we let the channel realization for each transmitter to be independently drawn from a given probability distribution. In the former case, we study both the atomic and non-atomic extensions of the game. In the atomic case, we show the existence of multiple NE and we use the best response dynamics to provide fully distributed algorithms to achieve a NE. This algorithm is proved to converge independently of the channel realization and the bandwidth allocated to each BS. We measure the price of anarchy of this solution and we observed that the performance of the fully decentralized solution (self-configured network) is close to that one obtained when there exists a central controller (optimally configured networks). In the non-atomic extension, we provide the optimal fractions of transmitter which must join each BS depending on their available bandwidths. Regarding, the BS sharing game, we show the existence of a unique NE. As in the previous case, we provide a fully decentralized algorithm which allows achieving a NE with probability one. Finally, we compare both scenarios and identify that BS selection performs better than BS sharing when there exists almost the same number or more transmitters than BSs. As identified in [4], this observation constitutes a Braess type paradox, which implies that increasing the space of strategies of each player, i.e., the

 

number of BSs each player can use, ends up degenerating the global performance of the network.

## II. System Model

*Notation:* In the sequel, matrices, vectors and scalars are respectively denoted by boldface upper case symbols, boldface lower case symbols, and italic lower case symbols. The transpose and Hermitian transpose of a vector $\boldsymbol{x}$ (matrix $\boldsymbol{X}$) is denoted by $\boldsymbol{x}^T$ and $\boldsymbol{x}^H$ (resp. $\boldsymbol{X}^T$ and $\boldsymbol{X}^H$). The sets of natural and real numbers are denoted by $\mathbb{N}$ and $\mathbb{R}$, respectively. Finite sets of natural numbers are denoted by calligraphic upper case letters. Given two sets denoted by $\mathcal{A}$ and $\mathcal{B}$, their Cartesian product is denoted by $\mathcal{A} \times \mathcal{B}$. The cardinality of set $\mathcal{A}$ is denoted by $|\mathcal{A}|$. The $S$-dimensional vectors $\boldsymbol{e}_s$, for all $s \in \{1, \dots, S\}$ and $S \in \mathbb{N}$, denotes a vector with zeros in all its entries except the $s$-th entry which is unitary. The operator $[x]^+$, with $x \in \mathbb{R}$, represents the operation $\max(0, x)$.

Consider a set $\mathcal{K} = \{1, \dots, K\}$ of transmitters and a set of $\mathcal{S} = \{1, \dots, S\}$ receivers, e.g., base stations (BS) or access points (AP). Each transmitter can access the network by using a (non-empty) set of BSs. Each BS operates in a specific frequency band and we neglect any type of interference due to the adjacent bands (adjacent channel interference). We denote the bandwidth associated with BS $s \in \mathcal{S}$ by $B_s$ and the total network bandwidth by $B = \sum_{s=1}^{S} B_s$. Each transmitter sends private messages only to its corresponding BSs and it does not exist any kind of information exchange between transmitters neither before nor during the whole transmission. Both transmitters and BSs are equipped with single antennas. Transmitter $k \in \mathcal{K}$ is able to simultaneously transmit to all the BSs subject to a power constraint,

$$\forall k \in \mathcal{K}, \quad \sum_{s=1}^{S} p_{k,s} \leqslant p_{k,\max}, \tag{1}$$

where $p_{k,s}$ and $p_{k,\max}$ respectively denote the transmit power dedicated to BS $s$ by transmitter $k$ and its maximum total transmit power. Without any loss of generality, we assume that all transmitters are limited by the same maximum transmit power level, i.e., $\forall k \in \mathcal{K}$, $p_{k,\max} = p_{\max}$.

For all $(k, s) \in \mathcal{K} \times \mathcal{S}$, we denote the channel coefficients between transmitter $k$ and BS $s$ by $h_{k,s}$. Each channel coefficient $h_{k,s}$ is a realization of a circularly symmetric complex Gaussian random variable $h$ with zero mean and unit variance. We consider a slow fading channel, so that all channel realizations remain constant during the transmission time. The baseband received signals sampled at symbol rate at BS $s$, denoted by $y_s$, can be written as a vector $\boldsymbol{y} = (y_1, \dots, y_S)^T$, such that

$$\forall s \in \mathcal{S}, \quad y_s = \boldsymbol{h}_s \boldsymbol{x}_s^T + \boldsymbol{w}. \tag{2}$$

Here, for all $(k, s) \in \mathcal{K} \times \mathcal{S}$, the $K$-dimensional vector $\boldsymbol{h}_s = (h_{1,s}, \dots, h_{K,s})$. The $K$-dimensional vector $\boldsymbol{x}_s = (x_{1,s}, \dots, x_{K,s})$, and $x_{k,s}$ represents the symbol sent by transmitter $k$ to BS $s$. The power allocation vector of transmitter $k$ is the vector $(p_{k,1}, \dots, p_{k,S})$, and $p_{k,s} = \mathbb{E}\left[x_{k,s} x_{k,s}^*\right]$ represents the power transmitted toward BS $s$ by player $k$. The $S$-dimensional vector $\boldsymbol{w} = (w_1, \dots, w_S)$, with $w_s \sim \mathcal{N}\left(0, \sigma_s^2\right)$ represents the noise at the receivers. Here, $\sigma_s^2 = N_0 B_s$, where $N_0$ denotes the noise spectral density.

The SINR of transmitter $k$ at BS $s$ is denoted by $\gamma_{k,s}$ and $\forall (k, s) \in \mathcal{K} \times \mathcal{S}$,

$$\gamma_{k,s} = \frac{p_{k,s} g_{k,s}}{\zeta_{k,s}}, \tag{3}$$

where, $\zeta_{k,s} = \sigma_s^2 + \displaystyle\sum_{j \in \mathcal{K} \setminus k} p_{j,s} g_{j,s}$ represents the noise plus multiple access interference (MAI) undergone by player $k$ at BS $s$ and $g_{k,s} = |h_{k,s}|^2$ represents the channel gains. We denote by $\mathcal{K}_s$ the set of transmitters using the BS $s$. Then, we define two different scenarios depending on the conditions over the sets $\mathcal{K}_s$ for all $s \in \mathcal{S}$. In the first scenario, named BS selection, each transmitter uses a unique BS. Thus, for all $s \in \mathcal{S}$, the sets $\mathcal{K}_s$ such that $|\mathcal{K}_s| > 0$ form a partition of the set $\mathcal{K}$, i.e., $\forall (j, k) \in \mathcal{S}^2$ and $j \neq k$, $\mathcal{K}_j \cap \mathcal{K}_k = \emptyset$ and $\mathcal{K}_1 \cup \ldots \cup \mathcal{K}_S = \mathcal{K}$. In the second scenario, named BS Selection, a given transmitter is allowed to simultaneously use several BSs. Thus, for all $s \in \mathcal{S}$, the sets $\mathcal{K}_s$ form a cover of the set $\mathcal{K}$, i.e., $\forall s \in \mathcal{S}, \quad \mathcal{K}_s \subseteq \mathcal{K}$. In the following two sections, we study both scenarios. Later, we compare their performance by simulation results.

### III. Base Station Selection Games

Assume that each transmitter can be modeled as a rational selfish player and that such an assumption is common knowledge among all players. Then, the BS selection scenario can be modeled by a non-cooperative game $\mathcal{G}_1$ described by the tuple $\left( \mathcal{K}, (\mathcal{P}_k)_{k \in \mathcal{K}}, (u_k)_{k \in \mathcal{K}} \right)$. Here, the set of transmitters $\mathcal{K}$ is the set of players. The strategy of a given player $k \in \mathcal{K}$ is its PA scheme, i.e., the $S$-dimensional PA vector $\boldsymbol{p}_k = (p_{k,1}, \ldots, p_{k,S}) \in \mathcal{P}_k$, where $\mathcal{P}_k$ is the set of all actions of player $k$ (strategy set). Since each player only transmits to a unique BS, its strategy set is defined as a finite set $\mathcal{P}_k$,

$$\begin{aligned}
\mathcal{P}_k &= \{ p_k \, \boldsymbol{e}_s : p_k \in [0, p_{\max}], \ \forall s \in \mathcal{S}, \ \boldsymbol{e}_s = (e_{s,1}, \ldots, e_{s,S}) \\
&\qquad \text{and} \ \forall r \in \mathcal{S} \setminus s, e_{s,r} = 0, \ \text{and} \ e_{s,s} = 1 \}.
\end{aligned} \tag{4}$$

Then, a strategy profile of the game is a super vector

$$\boldsymbol{p} = (\boldsymbol{p}_1, \ldots, \boldsymbol{p}_K) \in \mathcal{P},$$

where $\mathcal{P}$ is a finite set obtained from the Cartesian product of the strategy sets $\mathcal{P}_k$, for all $k \in \mathcal{K}$, i.e., $\mathcal{P} = \mathcal{P}_1 \times \ldots \times \mathcal{P}_K$. Let us denote by $\boldsymbol{p}_{-k}$ any vector in the finite set $\mathcal{P}_{-k} = \mathcal{P}_1 \times \ldots \times \mathcal{P}_{k-1} \times \mathcal{P}_{k+1} \times \ldots \times \mathcal{P}_K$. For a given $k \in \mathcal{K}$, the vector denoted by $\boldsymbol{p}_{-k}$ represents the strategies adopted by all the players other than player $k$. The utility function for player $k$, is defined as $u_k : \mathcal{P} \to \mathbb{R}_+$ and measures the satisfaction of player $k$ with respect to its chosen strategy [18]. In this study, we define the utility function for all players as their spectral efficiency, i.e., the ratio between their Shannon transmission rate and the available bandwidth $B$:

$$u_k(\boldsymbol{p}_k, \boldsymbol{p}_{-k},) = \sum_{s \in \mathcal{S}} \frac{B_s}{B} \log_2 \left( 1 + \gamma_{k,s} \right), \tag{5}$$

where $\gamma_{k,s}$ is given by (Eq. 3) and $\boldsymbol{p} \in \mathcal{P}$.

In the sequel, we consider a finite number of transmitters (players) such that each player is concerned with the strategy (BS selection and transmit power allocation) adopted by all the other players due to mutual interference.

We name this model: atomic BS selection game. In the second part, we consider a high number of players such that each of them is indifferent to the strategy adopted by every single player. In this case, each player is rather concerned with the fraction of players adopting the same strategy. We name this model non-atomic BS selection games.

### A. Atomic BS Selection Games

In the atomic extension of the BS selection game $\mathcal{G}_1$, our interest is to find a strategy profile $\boldsymbol{p}^* \in \mathcal{P}$ such that no player is interested in changing its own strategy. Once the network configuration $\boldsymbol{p}^*$ is reached, any unilateral deviation of a given player decreases its own utility. A network configuration $\boldsymbol{p}^*$ is known as a Nash equilibrium [19].

*Definition 1 (Nash Equilibrium):* In the strategic game $\mathcal{G}_1$, a strategy profile $\boldsymbol{p} \in \mathcal{P}$ is an NE if it satisfies, for all $k \in \mathcal{K}$ and for all $\boldsymbol{p}'_k \in \mathcal{P}_k$, that

$$u_k(\boldsymbol{p}_k, \boldsymbol{p}_{-k}) \geqslant u_k(\boldsymbol{p}'_k, \boldsymbol{p}_{-k}). \qquad (6)$$

In the following, we analyze the existence, multiplicity and determination of such strategy profiles.

*1) Existence of at least one NE:* Our first step toward identifying the strategy profiles leading to a NE is to prove that there exists at least one NE for any specific number of transmitters and BSs regardless of the channel realizations. There exist several methodologies for proving this [20]. In our case, we first show that the game $\mathcal{G}_1$ is a potential game (PG).

*Definition 2 (Exact Potential Game):* Any strategic game $\mathcal{G}$ defined by the tuple $\left( \mathcal{K}, (\mathcal{P}_k)_{k \in \mathcal{K}}, (u_k)_{k \in \mathcal{K}} \right)$ is an exact potential game (PG) if there exists a function $\phi(\boldsymbol{p})$ for all $\boldsymbol{p} \in \mathcal{P}$ such that for all players $k \in \mathcal{K}$ and for all $\boldsymbol{p}'_k \in \mathcal{P}_k$, it holds that

$$u_k(\boldsymbol{p}_k, \boldsymbol{p}_{-k}) - u_k(\boldsymbol{p}'_k, \boldsymbol{p}_{-k}) \quad = \quad \phi(\boldsymbol{p}) - \phi(\boldsymbol{p}'), \qquad (7)$$

where $\boldsymbol{p}' = (\boldsymbol{p}_1, \ldots, \boldsymbol{p}_{k-1}, \boldsymbol{p}'_k, \boldsymbol{p}_{k+1}, \ldots, \boldsymbol{p}_K)$.

Def. 2 together with Eq. (5) allow us to write the following proposition:

*Proposition 3:* The strategic game $\mathcal{G}_1$ is an exact potential game with potential function

$$\phi(\boldsymbol{p}) = \sum_{s \in \mathcal{S}} \frac{B_s}{B} \log_2 \left( \sigma_s^2 + \sum_{k=1}^{K} p_{k,s} g_{k,s} \right). \qquad (8)$$

Since the BS selection game $\mathcal{G}_1$ is a PG (Prop. 3), the following proposition (Prop. 4) is an immediate consequence of *Corollary 2.2* in [11].

*Proposition 4 (Existence of the NE):* The BS selection game $\mathcal{G}_1$ always has at least one NE in pure strategies.

*2) Multiplicity of the NE:* Once we have ensured the existence of at least one NE, we determine whether there exists a unique NE or several NE. As a first step, we show that rational players always transmit at the maximum power level $p_{\max}$:

*Proposition 5:* In the BS selection game $\mathcal{G}_1$, all players will always transmit at the maximum power independently of the channel chosen to transmit.

*Proof:* The utility function of player $k \in \mathcal{K}$ transmitting to a given BS $s \in \mathcal{S}$ is $u_k(p_k \boldsymbol{e}_s, \boldsymbol{p}_{-k}) = \log_2\left(1 + \frac{p_k g_{k,s}}{\zeta_{k,s}}\right)$. Then, since the logarithmic function is an increasing function, we have that $\forall (k,s) \in \mathcal{K} \times \mathcal{S}$, and $\forall p_k \in [0, p_{\max}]$, $u_k(p_k \boldsymbol{e}_s, \boldsymbol{p}_{-k}) = \log_2(1 + \frac{p_k g_{k,s}}{\zeta_{k,s}}) \leqslant u_k(p_{\max} \boldsymbol{e}_s, \boldsymbol{p}_{-k}) = \log_2(1 + \frac{p_{\max} g_{k,s}}{\zeta_{k,s}})$. Hence, rational players will always use their maximum transmit power level. ∎

Prop. 5 shows that the strategy set in (4) can be re-defined as follows

$$\begin{aligned} \mathcal{P}_k \quad = \quad &\{p_{\max} \, \boldsymbol{e}_s : \; \forall s \in \mathcal{S}, \; \boldsymbol{e}_s = (e_{s,1}, \ldots, e_{s,S}) \\ &\text{and} \; \forall r \in \mathcal{S} \setminus s, e_{s,r} = 0, \; \text{and} \; e_{s,s} = 1\}. \end{aligned} \tag{9}$$

The re-definition of the strategy sets $\mathcal{P}_k$ in Eq. (4) allows us to study the multiplicity of the NE by using basic elements of graph theory. First, let us index the elements of the strategy set $\mathcal{P}$ by using the set $\mathcal{I} = \{1, \ldots, S^K\}$ such that they are ordered following the index $i \in \mathcal{I}$. Denote by $\boldsymbol{p}^{(i)}$ the $i$-th element of the strategy set $\mathcal{P}$. Let us write each vector $\boldsymbol{p}^{(i)}$ with $i \in \mathcal{I}$, as a vector $\boldsymbol{p}^{(i)} = \left(\boldsymbol{p}_1^{(i)}, \ldots, \boldsymbol{p}_K^{(i)}\right)$, where for all $j \in \mathcal{K}$, $\boldsymbol{p}_j^{(i)} \in \mathcal{P}_j$. Second, consider that each of the strategy profiles $\boldsymbol{p}^{(i)}$ with $i \in \mathcal{I}$ is represented by a vertex $v_i$ in a given non-directed graph $G$. Each vertex is adjacent to the $K(S-1)$ vertices representing the strategy profiles obtained by letting only one player to change its own strategy. Let us denote by $\mathcal{V}_i$ the set of indices of the adjacent vertices of vertex $v_i$. More precisely, the graph $G$ can be defined by the tuple $G = (\mathcal{V}, \boldsymbol{A})$, where the set $\mathcal{V} = \{v_1, \ldots, v_{S^K}\}$ contains the $S^K$ possible strategy profiles of the game and $\boldsymbol{A}$ is a symmetric matrix (adjacency matrix of $G$) with dimensions $S^K \times S^K$ and entries defined as follows $\forall (i,j) \in \mathcal{I}^2$ and $i \neq j$,

$$a_{i,j} = a_{j,i} = \begin{cases} 1 & \text{if } i \in \mathcal{V}_j \\ 0 & \text{otherwise}, \end{cases} \tag{10}$$

and $a_{i,i} = 0$ for all $i \in \mathcal{I}$. In the non-directed graph $G$, we define the distance between vertices $v_i$ and $v_j$, for all $(i,j) \in \mathcal{I}^2$ as the length of the shortest path between $v_i$ and $v_j$. Considering the structure of $G$, a more precise definition can be formulated for the shortest path,

*Definition 6:* [Shortest Path] The distance (shortest path) between vertices $v_i$ and $v_j$, with $(i,j) \in \mathcal{I}^2$ in a given non-directed graph $G = (\mathcal{V}, A)$, denoted by $d_{i,j}(G) \in \mathbb{N}$ is:

$$d_{i,j}(G) = d_{j,i}(G) = \sum_{k=1}^K \mathbb{1}_{\left\{\boldsymbol{p}_k^{(i)} \neq \boldsymbol{p}_k^{(j)}\right\}}. \tag{11}$$

Note that the non-directed graph $G$ satisfies the property: $\forall (i,j) \in \mathcal{I}^2$, with $i \neq j$, $\quad 1 \leqslant d_{i,j}(G) \leqslant K$. Thus, for a specific number $S$ of BSs and $K$ transmitters, the maximum number of NE which can be observed is obtained as follows:

*Proposition 7:* In a given BS selection game $\mathcal{G}_1$ where the condition

$$\forall (i,j) \in \mathcal{I}, \; \text{with} \; i \neq j, \quad \phi\left(\boldsymbol{p}^{(i)}\right) \neq \phi\left(\boldsymbol{p}^{(j)}\right) \tag{12}$$

always holds, the maximum number of NE which can be observed is $S^{K-1}$.

*Proof:* Assume that a given strategy profile $\boldsymbol{p}^{(i)}$ (vertex $v_i$) with $i \in \mathcal{I}$ is a NE (Prop. 4). Then, given the condition (12) it follows that none of the vertices in the set $\mathcal{V}_i$ is a NE. Hence, two NE vertices must be separated

 

by a minimum distance two in the non-directed graph $G = (\mathcal{V}, \boldsymbol{A})$. Thus, we obtain the maximum number of NE by calculating the maximum number of vertices mutually separated by minimum distance two in $G$. Given any two vertices $v_i$ and $v_j$, for all $(i, j) \in \mathcal{I}^2$ with $i \neq j$ we have that $d_{i,j}(G) \geqslant 1$. Then, the vertex $v_i$ and all the vertices $v_j$ such that $j \in \mathcal{J}_{i,k} = \left\{ n \in \mathcal{I} \setminus \{i\} : \boldsymbol{p}_k^{(n)} \neq \boldsymbol{p}_k^{(i)} \right\}$, for any $k \in \mathcal{K}$, are separated by minimum distance $d_{i,j}(G) \geqslant 2$. Then, for any $(i, k) \in \mathcal{I} \times \mathcal{K}$, the set $\mathcal{J}_{i,k}$ has cardinality $|\mathcal{J}_{i,k}| = S^{K-1} - 1$. Then, the total number of points mutually separated by minimum distance 2 (including the reference vertex $v_i$) is $|\mathcal{J}_{i,k}| + 1 = S^{K-1}$, which completes the proof. ∎

*3) Determination of the NE:* To evaluate the number of NE of the game $\mathcal{G}_1$ for a specific set of channel gains, we use an oriented graph $\hat{G} = \left( \mathcal{V}, \hat{\boldsymbol{A}} \right)$, where the adjacency matrix $\hat{\boldsymbol{A}}$ is a non-symmetric square matrix whose entries are $\forall (i, j) \in \mathcal{I}^2$ and $i \neq j$,

$$\hat{a}_{i,j} = \begin{cases} 1 & \text{if} \quad i \in \mathcal{V}_j \text{ and } \phi\left(\boldsymbol{p}^{(j)}\right) > \phi\left(\boldsymbol{p}^{(i)}\right) \\ 0 & \text{otherwise ,} \end{cases} \tag{13}$$

and $a_{i,i} = 0$ for all $i \in \mathcal{I}$.

In the graph $\hat{G}$, we say that a given vertex $v_i$ is adjacent to vertex $v_j$, if and only if $\phi(\boldsymbol{p}_i) > \phi(\boldsymbol{p}_j)$ and $d_{ij}(G) = 1$. Note that the condition for adjacency in $\hat{G}$ represents the rationality assumption of players: A player changes its strategy if the new strategy brings a higher utility function, i.e., increases the potential function. In Fig. 1, we show an example of the non-directed $G$ and oriented $\hat{G}$ graphs for the case where $K = 3$ and $S = 2$.

From the definition of the matrix $\hat{\boldsymbol{A}}$, we have that a necessary and sufficient condition for a vertex $v_i$ to represent a NE strategy profile is to have a null out-degree: $\deg^+(v_i) = 0$ (sink vertex), in the oriented graph $\hat{G}$. Hence, obtaining the number of NE in the game $\mathcal{G}_1$ boils down to identifying all the sinks in the oriented graph $\hat{G}$. For doing so, it suffices to identify the indices of the rows of matrix $\hat{A}$ containing only zeros. If the $i$-th row of matrix $\hat{A}$ is a null vector, then the strategy profile $\boldsymbol{p}^{(i)}$ is a NE. However, building the matrix $\hat{A}$ requires complete CSI, since it is necessary to determine whether $\phi(\boldsymbol{p}^{(i)}) > \phi(\boldsymbol{p}^{(j)})$, $\phi(\boldsymbol{p}^{(i)}) = \phi(\boldsymbol{p}^{(j)})$ or $\phi(\boldsymbol{p}^{(i)}) < \phi(\boldsymbol{p}^{(j)})$ for all $i \in \mathcal{I}$ and $j \in \mathcal{V}_i$.

To determine a strategy profile leading to a NE, in a distributed fashion with a less restrictive CSI at each radio device, we introduce the following definition:

*Definition 8 (Random Walks):* A walk through an oriented graph $\hat{G}$ is an ordered list of vertices $v_{i_1}, \ldots, v_{i_N}$ such that vertex $v_{i_{n+1}}$ is adjacent to vertex $v_{i_n}$, with $i_n \in \mathcal{I}$ for all $n \in \{1, \ldots, N\}$, and $N \leqslant S^K$. We say that a walk is random if given a vertex $v_{i_n}$, the vertex $v_{i_{n+1}}$ is chosen randomly from the set $\mathcal{V}_{i_n}$.

From Def. 8, we have the following result:

*Proposition 9:* Any random walk in the oriented graph $\hat{G}$ ends in a vertex representing a NE.

*Proof:* Each step of the walk, i.e. the transition from vertex $v_{i_n}$ to $v_{i_{n+1}}$, can be interpreted as changing from one strategy profile $\boldsymbol{p}^{(i_n)}$, $i_n \in \mathcal{I}$ to another strategy profile $\boldsymbol{p}^{(i_{n+1})}$, $i_{n+1} \in \mathcal{V}_{i_n}$ such that $\phi(\boldsymbol{p}^{(i_n)}) < \phi(\boldsymbol{p}^{(i_{n+1})})$. Since there exists a finite number of vertices in the graph, it turns out that any sequence $\phi(\boldsymbol{p}^{(i_1)}) < \phi(\boldsymbol{p}^{(i_2)}) < \ldots < \phi(\boldsymbol{p}^{(i_N)})$, with $i_n \in \mathcal{I}$ for all $n \in \{1, \ldots, N\}$, is finite, i.e., $N \leqslant S^K$. Moreover, the walk is ended if and only if the vertex $v_N$ does not have any adjacent node, i.e., vertex $v_N$ is a sink vertex. From the definition of the adjacency matrix $\hat{A}$ in Eq. (13) it follows that any sink vertex represents a NE (Def. 1). Thus, any random walk in the oriented graph $\hat{G}$ ends in a NE. This completes the proof. ■

In practical terms, to perform a walk through the oriented graph $\hat{G}$ implies imposing certain rules on each transmitter of a given self-configuring network: (a) A given player changes its strategy if and only if the potential function can be strictly increased. (b) Two or more players do not change their strategy simultaneously. (c) All players have the same chances to update their strategies. The first condition derives from the fact that each player aims to maximize its own utility function. The second condition is to ensure that each step in the random walk is equivalent to going from a given vertex to one of its adjacent vertices. The third condition is to ensure a random walk, i.e., to ensure that each step is done with the same probability to any of the adjacent vertices. The last two conditions might require certain synchronization system among the transmitters.

---

**Algorithm 1** Base Station Selection Algorithm

---

**Require:** $\forall k \in \mathcal{K}$,

    MAI Vector: $\boldsymbol{\zeta}_k(0) = \left(\boldsymbol{\zeta}_{k,1}(0), \ldots, \boldsymbol{\zeta}_{k,S}(0)\right)$

    Channel Realizations: $\boldsymbol{g}_k = (g_{k,1}, \ldots, g_{k,S})$, $\forall k \in \mathcal{K}$

    $t \leftarrow 0$.

    **repeat**

        $t \leftarrow t + 1$

        **for** $k = 1$ to K **do**

            $s \leftarrow \underset{i \in \mathcal{S}}{\operatorname{argmax}} \ \log_2\left(p_{\max} g_{k,s} + \xi_{k,s}(t-1)\right)$

            $\boldsymbol{p}_k(t) \leftarrow p_{\max}\boldsymbol{e}_s$

            $\boldsymbol{\zeta}_k(t) \leftarrow \boldsymbol{\zeta}_k(t-1) + \left(\boldsymbol{p}_k(t) - \boldsymbol{p}_k(t-1)\right)\boldsymbol{g}_k^T$

    **until** $\boldsymbol{p}(t) = \boldsymbol{p}(t-1)$

---

Note that if the algorithm is implemented in a distributed way, each player $k \in \mathcal{K}$ requires the knowledge of two parameters. First, the MAI level at each BS, i.e., the vector $\boldsymbol{\zeta} = (\zeta_1, \ldots, \zeta_S)$, where $\zeta_s = \sigma_s^2 + \sum_{k \in \mathcal{K}} p_{k,s} g_{k,s}$ and which is common to all the players. Second, the channel realization with respect to each BS, i.e., the vector $\boldsymbol{g}_k = (g_{k,1}, \ldots, g_{k,S})$. Each element of the vector $\boldsymbol{\zeta}$ is obtained by feedback from the corresponding BS at a frequency higher than the reciprocal of the channel coherence time. Each element of the vector $\boldsymbol{g}_k$ must be estimated by transmitter $k$ using channel estimation techniques, e.g., combining channel reciprocity and training sequences in a two-way link. An important remark regarding the proposed algorithm is that the NE where a given walk ends, mainly depends on the starting vertex in the graph $\hat{G}$ and the order we let each player to update its strategy. Thus, if each player is randomly chosen for updating at a given point of time, it is not possible to predict the NE where a walk ends. Hence, this might lead us to the situation where the convergence point is a non-optimal NE regarding a global metric, e.g., the network spectral efficiency. We analyze the optimality issues in Sec. III-C. In Fig. 2 we show a walk through the directed graph $\hat{G}$ of a given BS selection game with $K = 5$ and $S = 2$ and a given set of channel realizations. The potential obtained at each possible strategy profile $\boldsymbol{p}^{(i)}$, i.e., $\phi\left(\boldsymbol{p}^{(i)}\right)$ with $i \in \mathcal{I}$ is plotted in Fig. 3 as a function of their index $i$. In Fig. 2, it can be seen how different walks end in different NE.

### B. Non-Atomic Base Station Selection Games

In the non-atomic BS selection game, we consider that there exists a large number of players, such that players are indifferent to the strategy adopted by any single player. Each player is rather concerned with simultaneous deviations of fractions of the total number of players. Let us denote by $x_s$ the fraction of players transmitting to BS $s$, and assume that

$$
\begin{aligned}
\forall s \in \mathcal{S}_k, \quad x_s &= \frac{|\mathcal{K}_s|}{K} \\
\sum_{i=1}^{S} x_i &= 1.
\end{aligned}
\tag{14}
$$

We denote the ratio between the available total bandwidth $B$ and the total number of transmitters $K$ by $\alpha = \frac{B}{K}$. Thus, the ratio between the available bandwidth at BS $s$ and $K$, denoted by $\alpha_s = \frac{B_s}{K}$, satisfies $\sum_{s=1}^{S} \alpha_s = \alpha$. Using, these notations, the potential function $\phi$ can be written as follows

$$
\begin{aligned}
\phi(\boldsymbol{p}) &= \sum_{s \in \mathcal{S}} \frac{B_s}{B} \log_2 \left( \sigma_s^2 + \sum_{k=1}^{K} p_{k,s} g_{k,s} \right) \\
&= \sum_{s \in \mathcal{S}} \frac{B_s}{B} \log_2 \left( \sigma_s^2 + p_{\max} \sum_{k \in \mathcal{K}_s} g_{k,s} \right) \\
&= \sum_{s \in \mathcal{S}} \frac{B_s}{B} \log_2 \left( K\, N_0\, \alpha_s + p_{\max} \underbrace{\frac{|\mathcal{K}_s|}{K}}_{x_s} \frac{K}{|\mathcal{K}_s|} \sum_{k \in \mathcal{K}_s} g_{k,s} \right) \\
&= \sum_{s \in \mathcal{S}} \frac{B_s}{B} \log_2 \left( N_0\, \alpha_s + (x_s p_{\max}) \frac{1}{|\mathcal{K}_s|} \sum_{k \in \mathcal{K}_s} g_{k,s} \right) \\
&\quad + \sum_{s \in \mathcal{S}} \frac{B_s}{B} \log_2(K)
\end{aligned}
\tag{15}
$$

Note that the term $\sum_{s \in \mathcal{S}} \frac{B_s}{B} \log_2(K)$ does not depend on the strategy of the players. Thus, following *Lemma 2.7* in [11], the function

$$
\tilde{\phi}(\boldsymbol{p}) = \sum_{s \in \mathcal{S}} \frac{B_s}{B} \log_2 \left( N_0\, \alpha_s + (x_s p_{\max}) \frac{1}{|\mathcal{K}_s|} \sum_{k \in \mathcal{K}_s} g_{k,s} \right)
\tag{16}
$$

can be considered as another exact potential function of the BS selection game $\mathcal{G}_1$. Now, we assume that the number of players grows toward infinity at the same rate that the bandwidth available at each BS, i.e.,

- $B \longrightarrow \infty$ and $K \longrightarrow \infty$,
- $\lim_{B,K \to \infty} \frac{B}{K} = \alpha < \infty$, and
- $\forall s \in \mathcal{S}, \lim_{B_s, K \to \infty} \frac{B_s}{K} = \alpha_s < \infty$.

From a practical point of view, when the number of transmitters grows toward infinity while the total bandwidth or number of BSs remain constant, the MAI becomes a dominant parameter and thus, independently of the strategy adopted by each player, their own utility function tends to zero. Thus, no quality of service can be guaranteed, for instance, in terms of minimum transmission rates. For avoiding such a situation, we have considered that the number of players grows to infinity at the same rate as the total bandwidth. This ensures that the utility function of each player does not tend to zero when the load (number of transmitters per BS) of the network is increased. Under these conditions, it holds that for all $s \in \mathcal{S}$, $|\mathcal{K}_s| \to \infty$, and thus:

$$\frac{1}{|\mathcal{K}_s|} \sum_{k \in \mathcal{K}_s} g_{k,s} \to \int_{-\infty}^{\infty} \lambda \mathrm{d} F_g(\lambda) = \Omega, \qquad (17)$$

where $F_g$ is the cumulative probability function associated with the probability density function $f_g$ of the random variable $g$ (channel gains) described in Sec. II: $\mathrm{d} F_g(\lambda) = f(\lambda) \mathrm{d} \lambda$.

This result allows us to write the function $\phi$ as a function of the fractions $x_1, \ldots, x_S$,

$$\tilde{\phi}(x_1, \ldots, x_S) = \sum_{s \in \mathcal{S}} \frac{\alpha_s}{\alpha} \log_2 \left( N_0 \, \alpha_s + x_s p_{\max} \Omega \right), \qquad (18)$$

and thus, finding a set of fractions such that no player is interested to modify, i.e. a NE in the non-atomic extension of the game $\mathcal{G}_1$ boils down to solve the following optimization problem (OP) [21],

$$\begin{cases} \max\limits_{x_1, \ldots, x_S \in \mathbb{R}_+} & \sum\limits_{s \in \mathcal{S}} \frac{\alpha_s}{\alpha} \log_2 \left( N_0 \, \alpha_s + x_s p_{\max} \Omega \right), \\ \text{s.t.} & \sum\limits_{i=1}^{S} x_i = 1 \ \text{ and } \ \forall i \in \mathcal{S}, \ x_i \geqslant 0, \end{cases} \qquad (19)$$

which has a unique solution of the form

$$\forall s \in \mathcal{S}, \quad x_s = \frac{B_s}{B}. \qquad (20)$$

In Fig. 4, we show the fractions $x_s$, with $s \in \mathcal{S}$, obtained by Monte-Carlo simulations and using Eq. (20) for a network with $S = 6$ BSs and $K = 100$ transmitters. Therein, it becomes evident that Eq. (20) is a precise estimation of the outcome of the non-atomic BS selection game. Note that if all the BSs are allocated with the same bandwidth $B_s = \frac{B}{S} \ \forall s \in \mathcal{S}$, the fraction of players at each BS is identical, i.e., $\forall s \in \mathcal{S}$, $x_s = \frac{1}{S}$. This result is a generalization of the one in [5], where similar fractions were obtained for the case where each BS is allocated with the same bandwidth and players observe the same channel gains, i.e., $\forall (k, s) \in \mathcal{K} \times \mathcal{S}$, $g_{k,s} = 1$.

### C. Efficiency of the Nash Equilibria

Here, we evaluate the performance of the network when a completely decentralized stable configuration is achieved (Nash equilibrium) and the performance when there exists a central controller that dictates a configuration which maximizes a given global metric. In this study, we consider as global metric, the sum of the utilities of each player, i.e., the network spectral efficiency. To carry out such a study we consider two metrics known in the game theory jargon as price of anarchy and price of stability [22]. We formally define such parameters as follows:

*Definition 10:* In the strategic game $\left(\mathcal{K}, (\mathcal{P}_k)_{k \in \mathcal{K}}, (u_k)_{k \in \mathcal{K}}\right)$, denote the set of NE strategy profiles by $\mathcal{P}^* \subseteq \mathcal{P}$. Then, the price of anarchy (PoA) and the price of stability (PoS) are the ratios,

$$\text{PoA} = \frac{\max\limits_{\boldsymbol{p} \in \mathcal{P}} \sum\limits_{k=1}^{K} u_k(\boldsymbol{p})}{\min\limits_{\boldsymbol{p} \in \mathcal{P}^*} \sum\limits_{k=1}^{K} u_k(\boldsymbol{p})}, \tag{21}$$

and

$$\text{PoS} = \frac{\max\limits_{\boldsymbol{p} \in \mathcal{P}} \sum\limits_{k=1}^{K} u_k(\boldsymbol{p})}{\max\limits_{\boldsymbol{p} \in \mathcal{P}^*} \sum\limits_{k=1}^{K} u_k(\boldsymbol{p})}, \tag{22}$$

respectively.

The discrete nature of the strategy set of the players makes obtaining a closed-form expression for both the PoA and PoS, a very difficult task. Thus, we evaluate both PoA and PoS using Monte-Carlo simulations. For instance, in Fig 5, we respectively plot the PoA and PoS for a network with $S \in \{2, 3\}$ BSs, and $K \in \{1, \ldots, 9\}$ players. Therein, we observe that the loss due to decentralization is minimum since $PoA \longrightarrow 1$, when the number of players increase. Similarly to the PoA, the PoS is also close to unity but different from the PoA. This implies that effectively, the case where several NE are observed, takes place often. As a concluding remark regarding the efficiency of NE, we state that the self-configuring nature of the network does not imply a significant loss of optimality, i.e., if the network were centralized by enforcing signaling protocols between all transmitters and the different BSs, the gain in network spectral efficiency will not justify the increment of signaling traffic due to the feedback of the optimal strategies.

### D. Equilibrium in Mixed Strategies

For any player $k \in \mathcal{K}$, let the vector $\boldsymbol{q}_k = (q_{k,1}, \ldots, q_{k,S})$ represent a discrete probability distribution over the set of pure strategies $\mathcal{P}_k$. Here, $q_{k,s}$ represents the probability of player $k$ transmitting to BS $s$. The mixed-strategy space of player $k$ is the standard simplex $\mathcal{Q}_k$:

$$\mathcal{Q}_k = \{(q_{k,1}, \ldots, q_{k,s}) \in \mathbb{R}^S : \sum_{s=1}^{S} q_{k,s} = 1,$$

$$\text{and } \forall s \in \mathcal{S}, \ q_{k,s} \geqslant 0\}, \tag{23}$$

and the space of mixed-strategies is $\mathcal{Q} = \mathcal{Q}_1 \times \ldots \times \mathcal{Q}_K$. Let $\boldsymbol{s} = (s_1, \ldots, s_K)$ be a vector in the discrete set $\mathcal{S}^K$. Let us also index each element of the set $\mathcal{S}^K$ with the set $\{1, \ldots, S^K\}$ such that elements are ordered following the index $i \in \{1, \ldots, S^K\}$. Denote by $\boldsymbol{s}^{(i)} = \left(s_1^{(i)}, \ldots, s_K^{(i)}\right)$, the $i$-th element of such a set $\mathcal{S}^K$. Denote by $\boldsymbol{p}_{-k}^{(i)}$, the vector

$$\boldsymbol{p}_{-k}^{(i)} = \left(\boldsymbol{p}_1^{(i)}, \ldots, \boldsymbol{p}_{k-1}^{(i)}, \boldsymbol{p}_{k+1}^{(i)}, \ldots, \boldsymbol{p}_K^{(i)}\right),$$

where for all $j \in \mathcal{K}$, $\boldsymbol{p}_j^{(i)} = p_{\max} \boldsymbol{e}_{s_j^{(i)}} \in \mathcal{P}_j$. In the mixed-strategy extension of the BS selection game $\mathcal{G}_1$, the utility function $U_k(\boldsymbol{q})$, with $\boldsymbol{q} = (\boldsymbol{q}_1, \ldots, \boldsymbol{q}_K) \in \mathcal{Q}$, is defined as the expected value of the corresponding pure strategy utilities with respect to the probability distributions $\boldsymbol{q}_k$ for all $k \in \mathcal{K}$, i.e.

$$U_k(\boldsymbol{q}_k, \boldsymbol{q}_{-k}) = \sum_{i=1}^{S^K} \prod_{k=1}^{K} q_{k, s_k^{(i)}} u_k(\boldsymbol{p}_k^{(i)}, \boldsymbol{p}_{-k}^{(i)}). \tag{24}$$

Following *Lemma 2.10* in [11], and since, the game $\mathcal{G}_1$ is an exact PG (Prop. 3), we claim the existence of a potential function in the mixed-strategy extension of $\mathcal{G}_1$. We denote such a potential by $\bar{\phi}(\boldsymbol{q})$,

$$\bar{\phi}(\boldsymbol{q}) = \sum_{i=1}^{S^K} \prod_{k=1}^{K} q_{k, s_k^{(i)}} \phi(\boldsymbol{p}^{(i)}), \tag{25}$$

where $\boldsymbol{q} \in \mathcal{Q}$. From [19], we know that there always exists a NE in mixed strategies for the game $\mathcal{G}_1$. Thus the following OP must have at least one solution,

$$\begin{cases} \max_{(q_{k,s})_{\forall (k,s) \in \mathcal{K} \times \mathcal{S}}} & \sum_{i=1}^{S^K} \prod_{k=1}^{K} q_{k, s_k^{(i)}} \phi(\boldsymbol{p}^{(i)}) \\ \text{s.t.} & \forall k \in \mathcal{K}, \quad \sum_{s=1}^{S} q_{s,k} = 1, \\ & \forall (k,s) \in \mathcal{K} \times \mathcal{S}, \quad q_{s,k} \geqslant 0. \end{cases} \tag{26}$$

However, the solution to the OP (26) might not be necessarily a fully mixed strategy, i.e., a vector $\boldsymbol{q}_k$ with more than one entry different from zero. Indeed, depending on the channel realizations, it is possible that no NE in fully mixed strategies is observed, i.e.,

$$\exists (k,s) \in \mathcal{K} \times \mathcal{S} : \forall \boldsymbol{q} \in \mathcal{Q}, \quad U_k(\boldsymbol{e}_s, \boldsymbol{q}_{-k}) > U_k(\boldsymbol{q}_k, \boldsymbol{q}_{-k}). \tag{27}$$

For instance, consider the case when there exist two players and two BSs, i.e., $K = 2$, and $S = 2$. Then, we obtain that if there exists a pair $(k,s) \in \mathcal{K} \times \mathcal{S}$, such that

$$\frac{g_{k,-s}}{g_{k,s}} \leqslant \frac{\sigma_{-s}^2}{\sigma_s^2 + p_{\max} g_{-k,s}}, \tag{28}$$

the condition (27) always holds and thus, it does not exist a NE in fully mixed strategies. Here, we denote by $-s$ and $-k$ the element other than $s$ and $k$, in the binary sets $\mathcal{S}$ and $\mathcal{K}$, respectively. Note that the non-existence of a NE in fully mixed strategies does not mean that it does not exist a NE in mixed strategies [19]. The existence of at least one NE in pure strategies has been proved (Prop. 4) and a pure strategy NE is also a (degenerated) mixed strategy NE.

There exists several algorithms to iteratively solve the OP (26). Those algorithms are known in the domain of machine learning theory as linear reward inaction and linear reward penalty [23]. Several applications of those algorithms are presented in [13], [24], [25]. Contrary to the algorithms presented in this paper, linear reward inaction and penalty algorithms requires to set up some parameters to refine the convergence speed and the accuracy of the obtained probability distributions [23]. These parameters depend on the channel realizations, which means that at each coherence time such parameters must be re-adjusted by training.

## IV. Base Station Sharing Games

In this section, we consider the case where each player can be associated with several BSs. Here, each player not only selects its set of BSs but also the specific power level to transmit to each of its BSs. We define this interaction as a strategic game denoted by $\mathcal{G}_2 = (\mathcal{K}, (\mathcal{P}_k)_{k \in \mathcal{K}}, (u_k)_{k \in \mathcal{K}})$, where the set $\mathcal{K}$ remains being the indices of each player as in the previous section, $\mathcal{P}$ represents the space of strategies, where $\mathcal{P} = \mathcal{P}_1 \times \ldots \mathcal{P}_K$ and for all $k \in \mathcal{K}$

$$
\begin{aligned}
\mathcal{P}_k \;=\; & \big\{ (p_{k,1}, \ldots, p_{k,S}) \in \mathbb{R}^S : \forall s \in \mathcal{S}, \; p_{k,s} \geqslant 0, \\
& \text{and } \sum_{s \in \mathcal{S}} p_{k,s} \leqslant p_{\max} \big\}.
\end{aligned}
$$

The utility function remains being the spectral efficiency of each player as defined by Eq. (5).

### A. Existence and Uniqueness of the NE

To study the NE of the BS Sharing game $\mathcal{G}_2$, we first introduce the following proposition:

*Proposition 11:* The BS sharing game $\mathcal{G}_2$ is an exact potential game with potential function $\phi(\boldsymbol{p})$ given by Eq. (8) for all $\boldsymbol{p} \in \mathcal{P}$.

Prop. 11 leads us to the following result:

*Proposition 12:* In the strategic game $\mathcal{G}_2$ the strategy profile $\boldsymbol{p}^* = (\boldsymbol{p}_1^*, \ldots, \boldsymbol{p}_K^*)$, with $\boldsymbol{p}_k^* = \left( p_{k,1}^*, \ldots, p_{k,S}^* \right)$, where for all $(k, s) \in \mathcal{K} \times \mathcal{S}$,

$$
p_{k,s}^* = \left[ \frac{B_s}{B\,\beta_k} - \frac{\zeta_{k,s}}{g_{k,s}} \right]^+ , \tag{29}
$$

is the unique NE of the game. The constant $\beta_k$ for each player $k$ is set to satisfy the condition $\sum_{s=1}^{S} p_{k,s} = p_{\max}$ and $\zeta_{k,s}$ represents the noise plus MAI overcome by player $k$ at BS $s$.

*Proof:* To prove the existence of at least one NE, we use the fact that the BS sharing game $\mathcal{G}_2$ is a PG (Prop. 11). Then, following *Corollary 2.2* in [11], the existence of at least one NE is ensured. Thus, proving the uniqueness of the NE ends up being equivalent to prove that the OP:

$$
\max_{\boldsymbol{p} \in \mathcal{P}} \quad \sum_{s \in \mathcal{S}} \frac{B_s}{B} \log_2 \left( \sigma_s^2 + \sum_{j \in \mathcal{S}_s} p_{j,s} g_{j,s} \right) \tag{30}
$$

posses a unique solution. Indeed, since the potential function $\phi$ is strictly concave on $\mathcal{P}$ and $\mathcal{P}$ is a simplex, and thus convex, the Karush-Khun-Tucker (KKT) conditions are necessary and sufficient conditions of optimality. Hence, we write:

$$
\begin{cases}
\forall (k, s) \in \mathcal{K} \times \mathcal{S}, & \dfrac{B_s}{B} \left( \dfrac{g_{k,s}}{p_{k,s} g_{k,s} + \zeta_{k,s}} \right) - \beta_k + \nu_k = 0 \\[2ex]
\forall k \in \mathcal{K}, & \beta_k \left( \displaystyle\sum_{s=1}^{S} p_{k,s} - p_{\max} \right) = 0 \\[2ex]
\forall (k, s) \in \mathcal{K} \times \mathcal{S}, & \nu_k p_{k,s} = 0,
\end{cases} \tag{31}
$$

The solution to the system of equations (31) is known to be unique and achieved by using the water-filling algorithm [2]. Such a solution is given by expression Eq. (29) where $\beta_k$ is uniquely determined to satisfy the condition $\sum_{s=1}^{S} p_{k,s} = p_{\max}$, for all $k \in \mathcal{K}$. This ends up the proof. ∎

*B. Determination of the NE*

The NE of the BS sharing game $\mathcal{G}_2$ is fully determined by Eq. (29). Here, we study a decentralized algorithm such that the NE in Prop. 12 can be achieved by players in a decentralized fashion. First, consider that the strategy space $\mathcal{P}$ is obtained from the Cartesian product of the closed convex sets $\mathcal{P}_k$, for all $k \in \mathcal{K}$. Second, note that the solution to the OP

$$\max_{x \in \mathcal{P}_k} \phi\left(\boldsymbol{p}_1, \ldots, \boldsymbol{p}_{k-1}, \boldsymbol{x}, \boldsymbol{p}_{k+1}, \ldots, \boldsymbol{p}_K\right) \tag{32}$$

for all $k \in \mathcal{K}$ and for all $\boldsymbol{p}_{-k} \in \mathcal{P}_{-k}$ is unique and can be determined by using the water-filling algorithm [2]. Thus, the OP (30) can be solved iteratively by using a non-linear Gauss-Seidel method. Denote by $\boldsymbol{p}_k(t)$ the solution to $\boldsymbol{p}_k$ at iteration $t$, where $t \in \mathbb{N}$ and $\boldsymbol{p}_k(t+1)$ is given by

$$\arg\max \phi\left(\boldsymbol{p}_1(t+1), \ldots, \boldsymbol{p}_{k-1}(t+1), \boldsymbol{p}_k(t), \boldsymbol{p}_{k+1}(t), \ldots, \boldsymbol{p}_K(t)\right).$$

Then, following Prop. 2.7.1 in [26], the convergence of the sequence $\{p_k(t)\}_{t \in \{1, \ldots, N\}}$ for $N >> 0$ is ensured. Based on this result, we introduce the algorithm Alg. 2 for the BS sharing game:

---

**Algorithm 2** Base Station Sharing Algorithm

---

**Require:** $\forall k \in \mathcal{K}$,

    MAI Vector: $\boldsymbol{\zeta}_k(0) = \left(\boldsymbol{\zeta}_{k,1}(0), \ldots, \boldsymbol{\zeta}_{k,S}(0)\right)$

    Channel Realizations: $\boldsymbol{g}_k = (g_{k,1}, \ldots, g_{k,S})$, $\forall k \in \mathcal{K}$

    $t \leftarrow 0$.

    **repeat**

        $t \leftarrow t + 1$

        **for** $k = 1$ to K **do**

        $\boldsymbol{p}_k(t) \leftarrow \underset{\boldsymbol{p}_k \in \mathcal{P}_k}{\arg\max} \sum_{s \in \mathcal{S}} \dfrac{B_s}{B} \log_2\left(p_{k,s} g_{k,s} + \xi_{k,s}(t-1)\right)$

        $\boldsymbol{\zeta}_k(t) \leftarrow \boldsymbol{\zeta}_k(t-1) + \left(\boldsymbol{p}_k(t) - \boldsymbol{p}_k(t-1)\right) \boldsymbol{g}_k^T$

    **until** $\boldsymbol{p}(t) = \boldsymbol{p}(t-1)$

---

In Fig. 6 we show the convergence to the maximum of the potential function $\phi$ using Alg. 2 for the case of a network with $K = 6$ transmitters and $S = 3$ BSs. Therein, we show both a round Robin and random updates. In both cases, the convergence is achieved in very few iterations.

## V. Performance Analysis

In this section, we use algorithms Alg. 1 and Alg. 2 to compare the global performance of the network when BS selection and BS sharing are used. We choose the network spectral efficiency as the global performance metric, i.e., the sum of the utilities of all players. In fig. 7, we plot the network spectral efficiency for a network with $S \in \{2, 4, 8\}$ BSs and $K \in \{2, \ldots, 60\}$ transmitters assuming an SNR of 10 dB for each player. We observe that when $K < S$ the BS sharing technique performs better than BS selection. However, when $K \geqslant S$, the performance of the BS selection is strongly superior to BS sharing. For a large number of transmitters both techniques perform similarly.

Note that the strategy space of each player is bigger in the BS sharing scenario. Thus, one can think that a better performance is always obtained by using BS sharing than using BS selection. Paradoxically, we have found that on the contrary, for nearly fully and fully loaded networks i.e., $K \simeq S$ and $K > S$, increasing the space of strategies of each player produces a global loss of performance. A similar paradox is observed in congestion games where adding extra capacity to the network ends up reducing the overall performance [17]. A similar paradox to the one presented in this work is also observed in [4], [27], [28], [29].

## VI. Concluding Remarks

We have investigated the BS selection and BS sharing scenarios in the context of self-configuring networks using a non-cooperative model focusing on the spectral efficiency of each transmitter. We have proved the existence of at least one NE in both cases. In the BS sharing game a unique NE is observed, whereas BS selection games might exhibit several. We have provided fully decentralized algorithms such that players can calculate their NE strategy based on local information and the MAI observed at each BS. We have observed that no significant gain would be achieved by introducing a central controller in the case of BS selection. The self-configured network performs almost identical to the optimally configured network.

Finally, we have identified that depending on the number of transmitters BS selection might perform better than BS sharing. This result implies a Braess type paradox, where increasing the strategy space of each players produces a degeneration of the global network spectral efficiency.

## VII. Acknowledgments

This work was partially supported by Alcatel-Lucent within the Alcatel-Lucent Chair in Flexible Radio at SUPELEC.

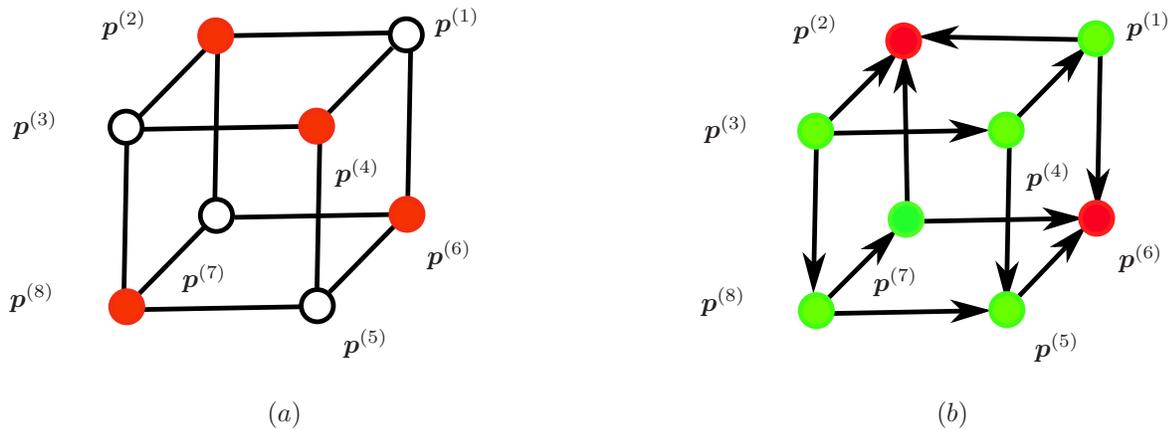

Fig. 1. (a) Non-oriented graph and (b) oriented graph representing the BS Selection game with $K = 3$, $S = 2$, under the condition $\phi(\boldsymbol{p}^{(2)}) > \phi(\boldsymbol{p}^{(6)}) > \phi(\boldsymbol{p}^{(1)}) > \phi(\boldsymbol{p}^{(5)}) > \phi(\boldsymbol{p}^{(4)}) > \phi(\boldsymbol{p}^{(7)}) > \phi(\boldsymbol{p}^{(8)}) > \phi(\boldsymbol{p}^{(3)})$. Total number of vertices: $S^K = 8$, number of neighbors per vertex: $K(S-1) = 3$. Maximum Number of NE: $S^{K-1} = 4$. Number of NE: 2 (red vertices in (b)).

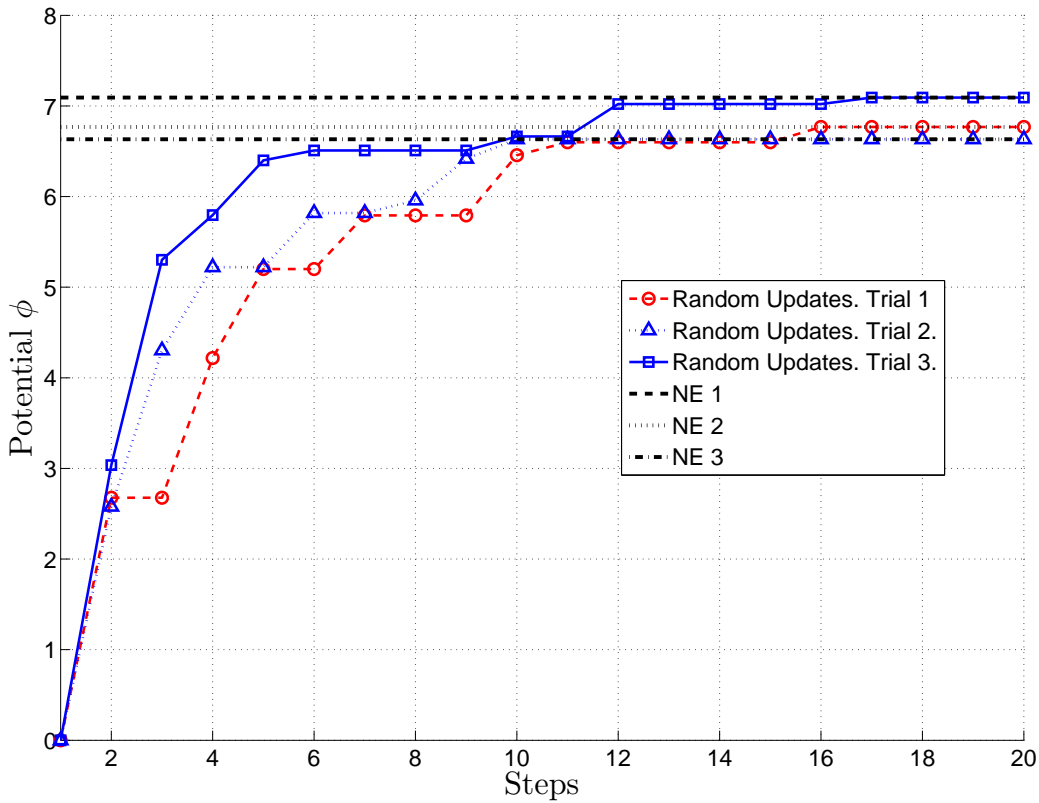

Fig. 2. Evolution of the potential at each update of the players using the BS Selection Alg. 1. At each step one player is randomly chosen to update its strategy. All the sequences are obtained using the same set of channel realizations $h_{k,s}$, $\forall (k,s) \in \mathcal{K} \times \mathcal{S}$. Number of players $K = 5$, Number of BSs $S = 3$, $\frac{B_1}{B} = 0.14$, $\frac{B_2}{B} = 0.40$, and $\frac{B_3}{B} = 0.46$. $SNR = 10 \log_{10}\left(\frac{p_{\max}}{N_0 B}\right) = 10$ dB.

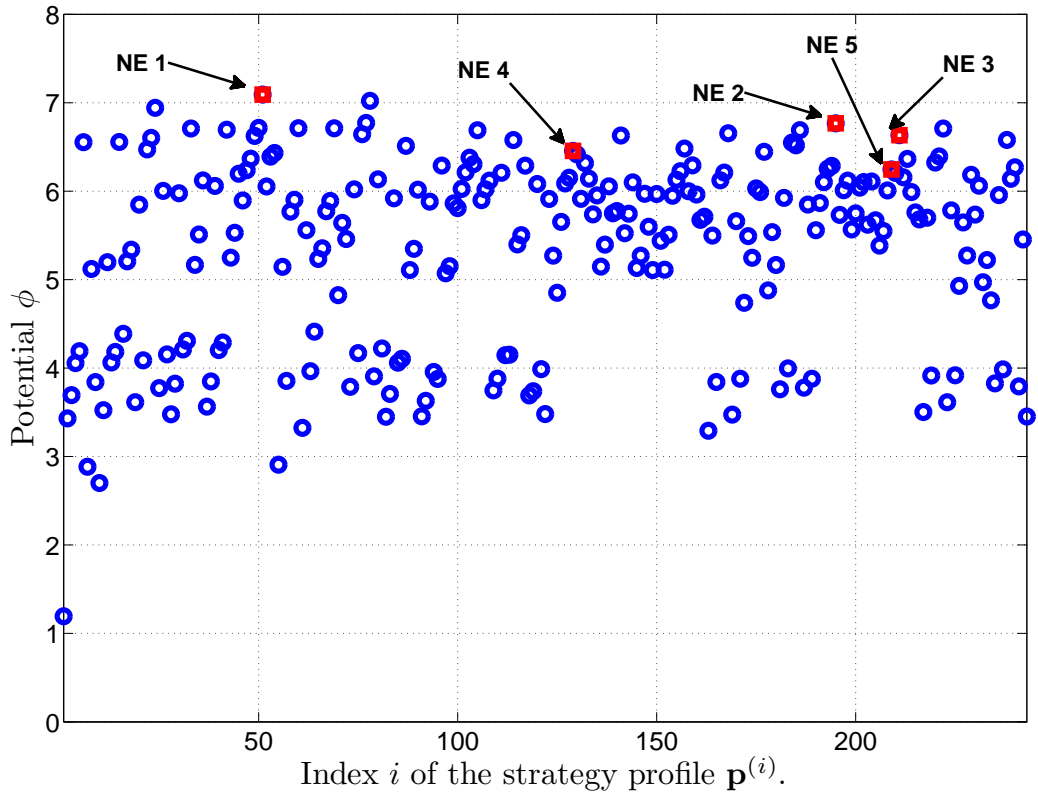

Fig. 3. Potential associated with each strategy profile $\boldsymbol{p}^{(i)}$ as a function of $i \in \mathcal{I}$. The set of channel realizations $h_{k,s}$, $\forall (k,s) \in \mathcal{K} \times S$ is identical to that one used in Fig. 2. Nash Equilibria are pointed by arrows. Number of players $K = 5$, Number of BSs $S = 3$, $\frac{B_1}{B} = 0.14$, $\frac{B_2}{B} = 0.40$, and $\frac{B_3}{B} = 0.46$. $SNR = 10 \log_{10} \left( \frac{p_{\max}}{N_0 B} \right) = 10$ dB.

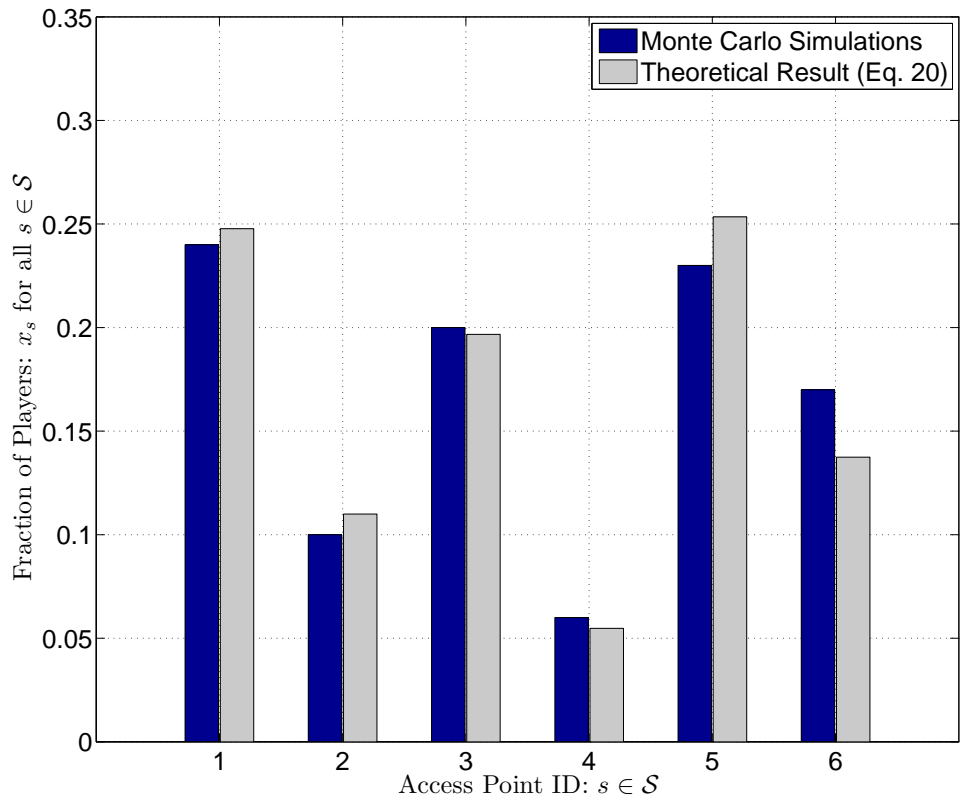

Fig. 4. Fraction of players transmitting to BS $s$, with $s \in \mathcal{S}$, calculated using Monte-Carlo simulations and using Eq. (20) for a network with $S = 6$ BSs and $K = 100$ transmitters, with $\boldsymbol{\alpha} = (\alpha_s)_{\forall s \in \mathcal{S}} = (0.25, 0.11, 0.20, 0.05, 0.25, 0.14)$. $SNR = 10\log_{10}\left(\frac{p_{\max}}{N_0 B}\right) = 10$ dB.

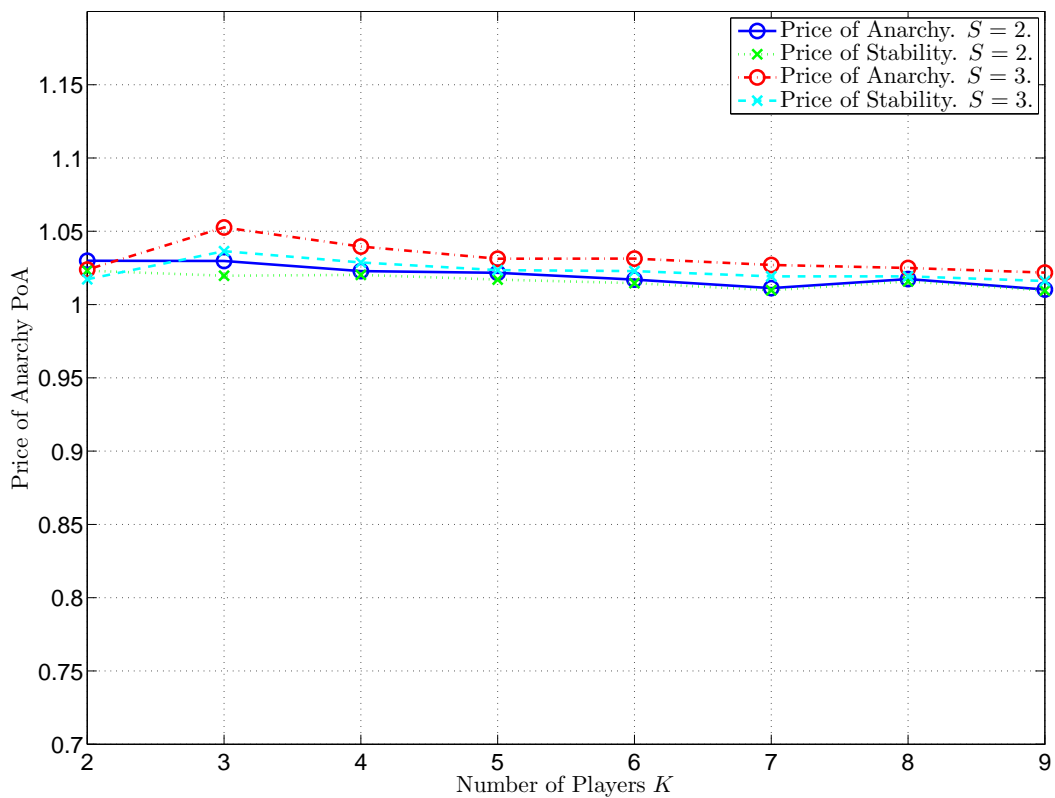

Fig. 5. Price of anarchy (Def. 10) and price of stability (Def. 10)as a function of the number of players $K$ for the case of two and three BSs, $S = 2$ and $S = 3$. $SNR = 10 \log_{10} \left( \frac{p_{\max}}{N_0 B} \right) = 10$ dB.

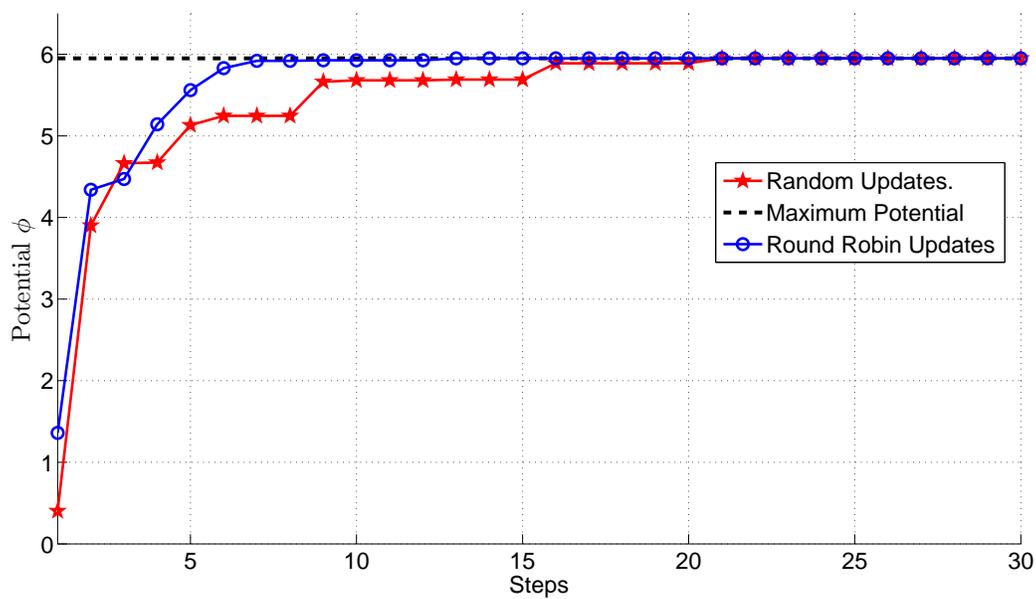

Fig. 6. Evolution of the potential at each update of the players using the BS Sharing Alg. 2. Number of players $K = 6$, Number of BSs $S = 3$, $\frac{B_1}{B} = 0.75$, $\frac{B_2}{B} = 0.21$, and $\frac{B_3}{B} = 0.04$. $SNR = 10 \log_{10}\left(\frac{p_{\max}}{N_0 B}\right) = 10$ dB.

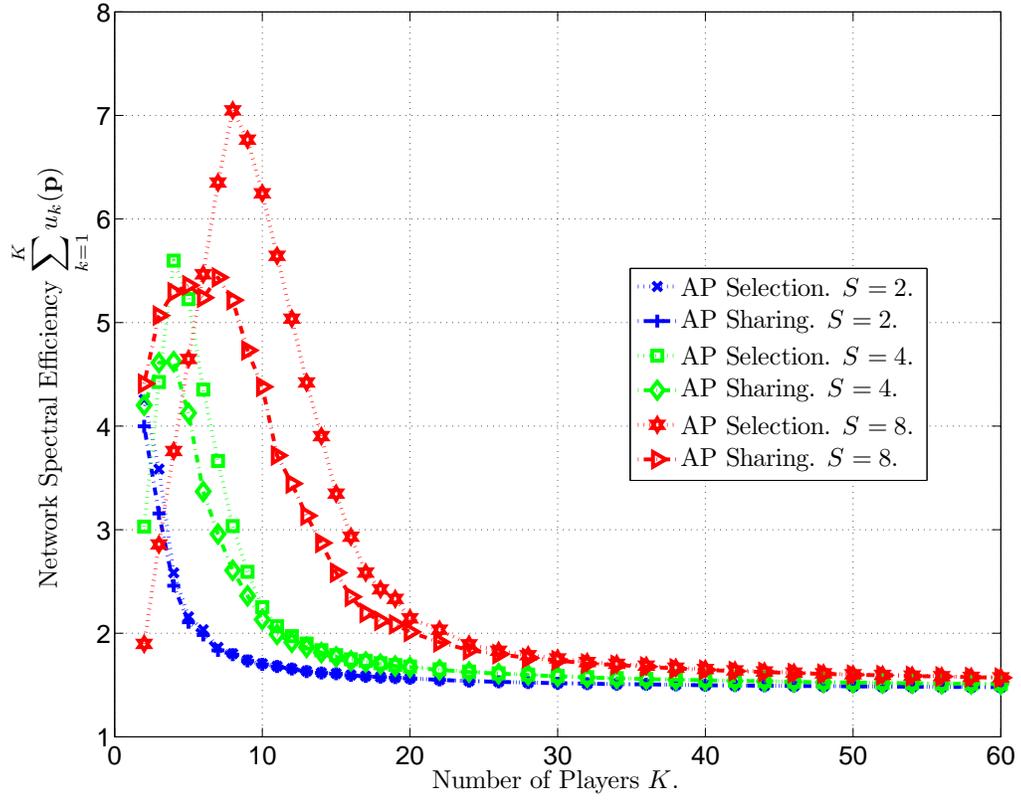

Fig. 7.   Sum spectral efficiency for the BS sharing and BS Selection algorithms as a function of the number of players present in the network. Players operate at a transmit SNR $10 \log_{10} \left( \frac{p_{\max}}{N_0 B} \right) = 10$ dB and the fractions $\frac{B_s}{B} = \frac{1}{S}$, $\forall s \in \mathcal{S}$. $SNR = 10 \log_{10} \left( \frac{p_{\max}}{N_0 B} \right) = 10$ dB.